\documentclass[journal=ancac3,manuscript=article,layout=twocolumn]{achemso}

\usepackage{chemformula} 
\usepackage[T1]{fontenc} 

\usepackage[separate-uncertainty=true]{siunitx}
\usepackage{bpchem}
\usepackage{romannum}
\usepackage{import}
\usepackage{transparent}
\usepackage{amsmath}

\DeclareSIUnit{\molar}{M}

\author{Kristin E. J. Kühl}
\affiliation{Department of Physics and Research Center OPTIMAS, RPTU University Kaiserslautern-Landau, 67663 Kaiserslautern, Germany}
\email{kuehlk@rptu.de}

\author{Katharina Rediger}
\affiliation{Department of Chemistry, RPTU University Kaiserslautern-Landau, 67663 Kaiserslautern, Germany}

\author{Nikhita Khera}
\affiliation{Department of Physics and Research Center OPTIMAS, RPTU University Kaiserslautern-Landau, 67663 Kaiserslautern, Germany}

\author{Ephraim Spindler}
\affiliation{Department of Physics and Research Center OPTIMAS, RPTU University Kaiserslautern-Landau, 67663 Kaiserslautern, Germany}

\author{Gereon Niedner-Schatteburg}
\affiliation{Department of Chemistry, RPTU University Kaiserslautern-Landau, 67663 Kaiserslautern, Germany}

\author{Elke Neu}
\affiliation{Department of Physics and Research Center OPTIMAS, RPTU University Kaiserslautern-Landau, 67663 Kaiserslautern, Germany}

\author{Mathias Weiler}
\affiliation{Department of Physics and Research Center OPTIMAS, RPTU University Kaiserslautern-Landau, 67663 Kaiserslautern, Germany}

\author{Georg von Freymann}
\affiliation{Department of Physics and Research Center OPTIMAS, RPTU University Kaiserslautern-Landau, 67663 Kaiserslautern, Germany}
\alsoaffiliation{Fraunhofer Institute for Industrial Mathematics ITWM, 67663 Kaiserslautern, Germany}

\title[]
  {Direct Laser Writing of Ferromagnetic Nickel Utilizing the Principle of Sensitized Triplet-Triplet Annihilation Upconversion}

\abbreviations{\begin{description}
    \item[2PA] two-photon absorption 
    \item[DIPEA] diisopropylethylamine
    \item[DLW] Direct Laser Writing
    \item[DMF] 2,5-dimethylfuran
    \item[DMI] 1,3-Dimethyl-2-imidazolidinone
    \item[EDX] energy dispersive X-ray spectroscopy
    \item[ESR] electron spin resonance
    \item[NV] nitrogen vacancy
    \item[ODMR] optically detected magnetic resonance
    \item[PTOEP] platinum octaethylporphyrin
    \item[SCE] saturated calomel electrode 
    \item[sTTA-UC] sensitized triplet-triplet annihilation upconversion
    \item[TTET] triplet triplet energy transfer  
\end{description}}


\begin{document}







\begin{abstract}
Direct laser writing of ferromagnetic microstructures is of great interest for sensing and data storage in compact three-dimensional architectures. However, reliable direct laser writing of metallic and even more so ferromagnetic materials remains a major challenge. Here, we present a novel photoresist suitable to direct laser write ferromagnetic nickel based on sensitized triplet–triplet annihilation upconversion. By combining an in-situ photochemical deoxygenation process with a sensitized triplet–triplet annihilation upconversion process as well as a photoreduction of \ch{Ni^{2+}} ions, the deposition of metallic nickel is enabled under ambient conditions. Using this approach, nickel structures are fabricated as a proof of concept. Scanning electron microscopy and EDX analysis confirm the spatially confined deposition of nickel, while magnetic characterization by vibrating sample magnetometry and scanning NV magnetometry demonstrate the ferromagnetic nature of the printed structures. This work presents a major step forward in extending the possibilities of direct laser writing to metallic and ferromagnetic materials.
\end{abstract}

\section{Introduction}

Complex three-dimensional microstructures are attracting increasing interest in science and technology and promise the realization of a wide variety of applications extending from the field of 
plasmonics\cite{schuller2010plasmonics,yang2017wavevector,anker2008biosensing,hu2016ultrafast} through aviation and automotive industries\cite{gibson2021additive,thompson2016design} to biotechnology
and medicine\cite{thompson2016design,otuka2021two,soto2020medical}. While applications so far rely mainly on polymeric and ceramic hybrid materials, manufacturing such structures from metals and especially from non-noble metals, continues to challenge scientists.

Direct laser writing (DLW) is a versatile method for additive manufacturing of microscale polymeric structures \cite{maruo1997three, skliutas2025multiphoton}. Until now, there exists only a limited number of proposals to fabricate metallic structures using DLW. In 2006 Tanaka \textit{et al.} reported fabrication of gold and silver structures from aqueous solutions of \ch{HAuCl_4} and \ch{AgNO_3} \cite{tanaka2006two}. Over the course of time structural quality improved: Blasco \textit{et al.} fabricated conductive gold-containing composite structures by simultaneous photopolymerization of an acrylate-functionalized poly(ethylene glycol) derivative and photoreduction of \ch{HAuCl_4} \cite{blasco2016fabrication} and Waller \textit{et al.} demonstrated highly conductive silver structures from a gelatin containing aqueous solution of \ch{AgClO_4} \cite{waller2021photosensitive}. However, those methods are focused on noble metals, which are useful for highly conducting structures requested, \textit{e.g.} in plasmonic applications. Applications like microelectronic devices such as sensors and actuators \cite{zhang20253d} or fully motion-controlled microrobots \cite{zhang2025biohybrid} would benefit from the possibility of printing ferromagnetic materials. However, due to their lower standard potentials, which corresponds to a smaller thermodynamic driving force for electron uptake\cite{harris2014lehrbuch}, such non-noble metals are more difficult to deposit from solutions, which calls for more complex processes to enable manufacturing using DLW.

Typically, DLW utilizes a two-photon absorption (2PA) process to initiate the photoreaction inside a very small volume defined by the isointensity surface of a tightly focused laser beam. This volume is often called voxel, short for volume pixel, and often has a lateral diameter lower than \SI{100}{\nano\meter} and an axial extension of approximately \SI{300}{\nano\meter}, depending on parameters such as laser power\cite{hohmann2015three}. The photoreaction is confined to this volume only due to a nonlinear (mostly quadratic) relationship between laser intensity and the polymerization threshold \cite{maruo1997three,skliutas2025multiphoton}. While 2PA inherently provides this nonlinearity, current research also addresses other multiphoton processes to provide a nonlinearity for the printing process. Hahn \textit{et al.} used consecutive two-step absorption instead of simultaneous 2PA \cite{hahn2021two}. Awwad \textit{et al.} and Z. Wang \textit{et al.} induced polymerization \textit{via} sensitized triplet-triplet annihilation upconversion (sTTA-UC)\cite{awwad2020visible,wang2022three}. This approach exploits the conversion of two low-energy excitations into a high-energy excitation. In this way, the use of expensive femtosecond pulsed lasers can be avoided while maintaining a quadratic dependence of the concentration of the printing-relevant excited states on the incident light intensity.

Y. Wang \textit{et al.} used two-photon decomposition of metal carbonyls in combination with optical force trapping and sintering with the printing laser to print molybdenum, cobalt and tungsten structures. This trapping and sintering process is enhanced by the localized surface plasmon resonance the laser excites in the metal nanoparticles\cite{wang2024free}. With this technique, they achieved writing speeds of up to \SI{20}{\micro\meter\per\second} \cite{wang2024free}. However, their resist needs to be prepared in a glove box under inert atmosphere and sealed in a liquid cell to shield it from oxygen, which would quench the process. 

In this work, we present a new approach to DLW of nickel utilizing a sTTA-UC process to initiate the photoreduction of nickel ions by a sacrificial electron donor in combination with \textit{in situ} deoxygenation. This introduces active protection of triplet states by incorporating an oxygen scavenger within the system\cite{baluschev2016annihilation}, which increases the writing speed up to \SI{100}{\micro\meter\per\second}.

\section{Results and Discussion}
\subsection{Underlying photochemical reactions}

\begin{figure}[ht]         
	\centering                
	\def\svgwidth{250pt}    
	\includegraphics[width = 0.5\textwidth]{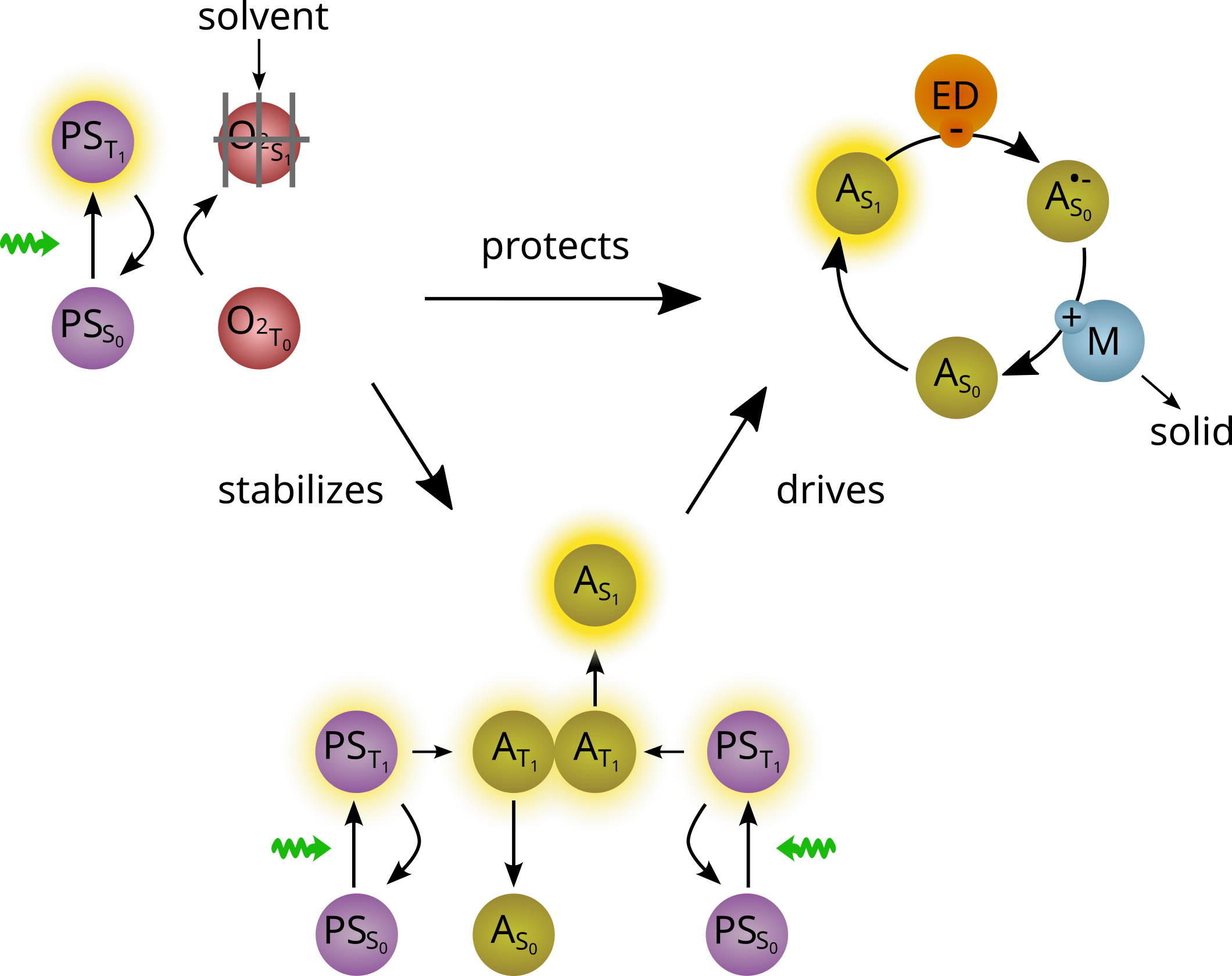}
	\caption{\small{Graphical representation of the cooperating chemical processes. The fundamental process is the photochemical deoxygenation performed by excitation of the photosensitizer (PS), generating singlet state oxygen (${\text{O}_{2_{\text{S}_1}}}$) by energy transfer. The excited state of molecules is symbolized by a yellow halo. This ${\text{O}_{2_{\text{S}_1}}}$ is permanently removed by oxidation of the solvent, symbolized by bars (left part). This provides a local deoxygenated volume which enables a sTTA-UC process generating the annihilator (A) in its first excited singlet state \ch{S_1}. The annihilator \ch{S_1} enables the photoreduction of metal ions into neutral atoms that precipate into bulk metal (M) \textit{via} electron transfer by an electron donor (ED) (right part).}}    
	\label{fig:idea_processes}    
\end{figure}

We combine three photochemical processes, a graphical representation of which is displayed in Fig. \ref{fig:idea_processes}. All processes are running in parallel. In a first process, the surrounding solvent is locally deoxygenated: the photosensitizer singlet $\text{PS}_{\text{S}_0}$ is excited into its triplet state $\text{PS}_{\text{T}_1}$, which is quenched through energy transfer to the dissolved ${\text{O}_{2_{\text{T}_0}}}$, generating singlet state oxygen ${\text{O}_{2_{\text{S}_1}}}$. ${\text{O}_{2_{\text{S}_1}}}$ molecules are then permanently removed from the solution by a reaction with the molecules of the surrounding solvent that serves as oxygen scavenger. These locally oxygen-free conditions enable a second process: sTTA-UC excites the annihilator to its singlet state $\text{A}_{\text{S}_1}$ required for following the redox reaction. 
In this third process, the annihilator in the \ch{S_1} state generated by sTTA-UC is quenched by electron transfer from a sacrificial electron donor (ED) to the annihilator. The thus reduced annihilator proliferates its extra electron to the nickel cation. As an alternative pathway, the excited annihilator $\text{A}_{\text{S}_1}$ could transfer energy to the nickel cation, which is then subsequently reduced by the electron donor. Either of both pathways would need to take place twice in order to generate \ch{nickel^{(0)}}. 

\begin{figure*}         
	\centering     
	\fontsize{7pt}{12pt}\selectfont            
	\def\svgwidth{280pt}    
 	\includegraphics[width = 0.8\textwidth]{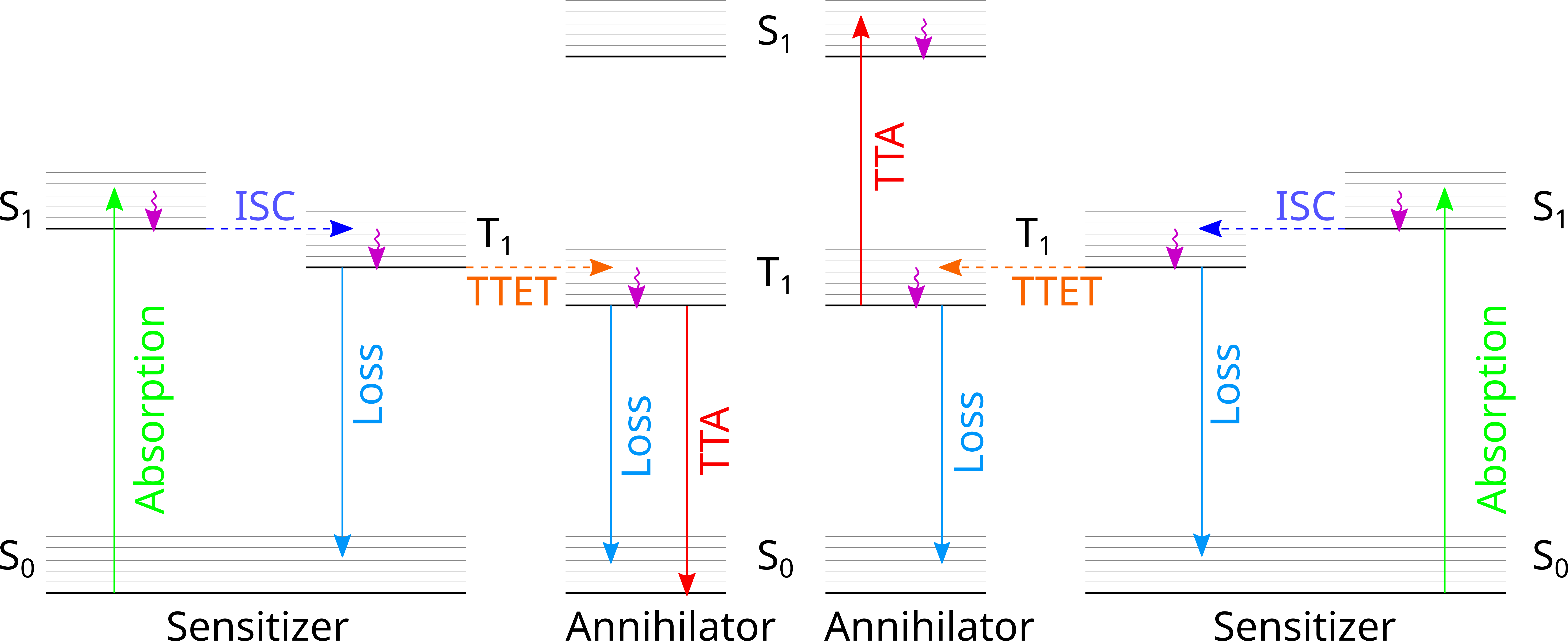}
	\caption{\small{Energy level diagram showing the steps of the triplet-triplet annihilation upconversion process. First, light gets absorbed by the sensitizer (green arrow). After internal conversion and vibrational relaxation (pink arrow) to its first singlet state, the triplet state $\text{T}_1$ is formed \textit{via} intersystem crossing (dashed dark blue arrow). Next, TTET occurs between the emitter and sensitizer (dashed orange arrow). When two annihilator molecules in the $\text{T}_1$ state collide, triplet-triplet annihilation upconversion occurs exciting one molecule to a higher energy singlet state with the second relaxing to the ground state (red arrows). The bright blue arrows represent potential non-radiative decay during the process of sTTA-UC. Adapted from \cite{bennison2021organic}.}}   
\label{fig:sTTA-UC}         
\end{figure*}

As sTTA-UC is not very well established in the DLW community, we depict the principle in Fig. \ref{fig:sTTA-UC} in greater detail. In the first step, a 532 nm continuous wave laser excites a photosensitizer to its first excited singlet state $\text{S}_1$, which is converted to the $\text{T}_1$ triplet state \textit{via} intersystem crossing. This is followed by a triplet-triplet energy transfer (TTET) from the triplet state of the photosensitizer to the triplet state of the annihilator. Two annihilator molecules in the $\mathrm{T}_1$
state may undergo triplet-triplet annihilation by an upconversion process that transforms one of the two in the ground state $\text{S}_0$ and one in the excited state $\text{S}_1$\cite{huang2020highly,edhborg2022best}. Due to the reaction of two sensitizer molecules, sTTA-UC realizes the required nonlinearity for the DLW process\cite{wang2022three}. 

\subsection{Implementation of the specified processes in the resist} 

\begin{table}
    \centering
    \begin{tabular}{c|c}
        Compound & Function in the resist \\ \hline
        Erythrosine B & photosensitizer\\ 
        perylene &  annihilator\\
        DIPEA & sacrificial electron donor\\
        $\mathrm{NiCl}_2\cdot 6\mathrm{H}_2\mathrm{O}$ &  supplier of $\mathrm{nickel}^{2+}$ cations \\
        DMI & deoxygenating solvent\\
    \end{tabular}
    \caption{The employed compounds and their proposed function}
    \label{table: compounds and functions}
\end{table}

After this schematic description of the underlying mechanism, we will now investigate the experimental realization of the various processes. The used compounds and their proposed functions in the resist can be found in Table \ref{table: compounds and functions}. Various studies have shown that efficient sTTA-UC processes are usually limited to oxygen-free environments since the triplet ground state of molecular oxygen acts as a quencher for the employed triplet states \cite{simon2012low,yanai2016recent}. However, efficient sTTA-UC in aerated solutions is achieved by deoxygenating solvents which are binding molecular singlet oxygen, which is formed from the triplet state by the excited sensitizer in the first process \cite{wan2018photochemically,lutkus2019singlet}. Here, 1,3-Dimethyl-2-imidazolidinone (DMI) is used for this purpose \cite{wan2018photochemically} and the deoxygenating properties are verified using the oxygen sensitive phosphorescence of platinum octaethylporphyrin (PTOEP) as established by Wan \textit{et al.} \cite{wan2018photochemically} before. Upon continuous irradiation ($\lambda = \SI{532}{\nano\meter}$) the integrated phosphorescence intensity of a solution of \SI{10e-5}{M} PTOEP in DMI increased by more than 43-fold within \SI{5}{\min} of irradiation time (Fig. S1). UV/vis absorption spectra recorded before and after irradiation exhibited no observable changes, indicating that the concentration of PTOEP was unaffected by the irradiation. 
Additionally, the excited state lifetime increased from \SI{670(6)}{\nano\second} (Fig. S2) before excitation to \SI{1746(13)}{\nano\second} (Fig. S3) after \SI{5}{\min}, confirming the decreasing oxygen concentration in solution further. Thus, DMI was chosen as a deoxygenating solvent for the photoresist. The sensitizer 2-(6-hydroxy-2,4,5,7-tetraiodo-3-oxoxanthen-9-yl)benzoic acid (Erythrosine B) used in the sTTA-UC process also serves as a sensitizer for generating singlet oxygen\cite{jiang2023production}.

Due to the \textit{in situ} establishment of an oxygen-free local environment, the sTTA-UC process is enabled to operate efficiently. For this purpose, Erythrosine B was used as sensitizer, since the energy of the donating $\text{T}_1$ state ($\text{E}_{{\text{T}_1}}\text{(Erythrosine B)} = \SI{1.89}{\electronvolt}$)\cite{tran1993intramolecular} is sufficiently higher than the $\text{T}_1$ energy of the used annihilator perylene (per) ($\text{E}_{{\text{T}_1}}\text{(per)} = \SI{1.53}{\electronvolt})$\cite{wu2012light}, thus enabling an efficient TTET to take place\cite{huang2020highly,edhborg2022best}. Also, the excited singlet state of perylene ($E_{{\text{S}_1}}\text{(per)} = \SI{2.76}{\electronvolt}$)\cite{wu2012light} can be populated upon recombining two triplet molecules $\text{E}_{{\text{T}_1}}\text{(per)}$, which is a fundamental requirement for a triplet-triplet annihilation upconversion process\cite{huang2020highly,edhborg2022best}. Thus, we check Erythrosine B and 100 eq. perylene in DMI mixture for luminescence of the upconverted $\text{S}_1$ state and weak upconverted emission was found with a maximum at \SI{474}{\nano\meter} (Fig. S4). However, it is suggested that the intensity is predominantly attenuated by reabsorption effects of the photosensitizer in the solution. The perylene molecule is in its excited $\text{S}_1$ state once the sTTA-UC takes place, from where it is able to start the photocatalytic reduction of nickel. The Erythrosine B molecule relaxes back to its ground state, ready to act as a photosensitizer again.

\begin{figure}       
	\centering                 
	\def\svgwidth{830pt}    
 	\includegraphics[width = 0.47\textwidth]{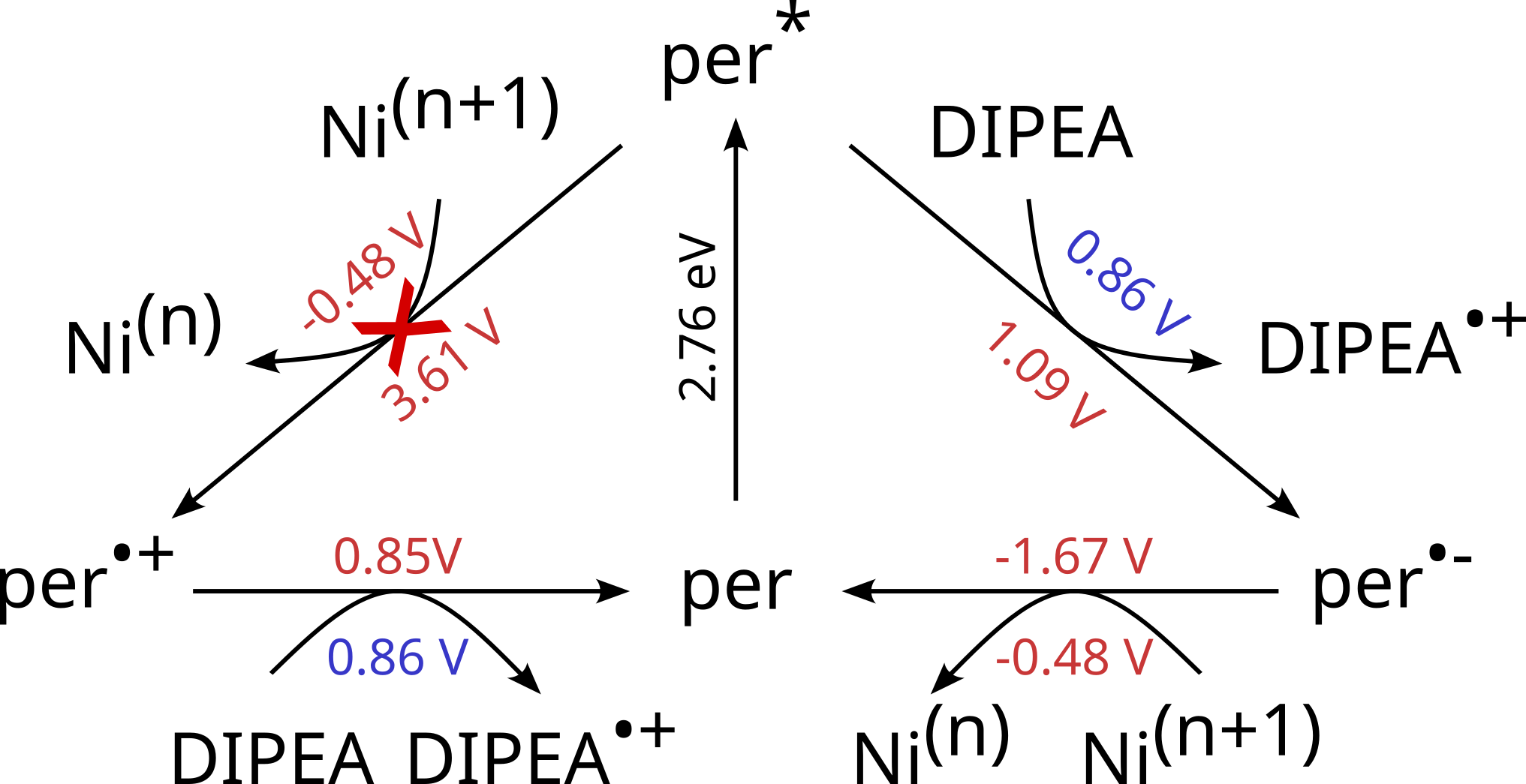}
	\caption{\small Latimer diagram of a photocatalytic reduction of nickel ions by perylene (per) with DIPEA as an electron donor. The central vertical arrow indicates excitation of perylene to the $\mathrm{S}_1$ state with the corresponding excitation energy. Apart from this one, each arrow represents a redox half-reaction between two adjacent oxidation states. Red numbers indicate reduction and blue numbers oxidation potentials, both referenced to SCE. Electron transfer is thermodynamically favorable when the reduction potential of the acceptor exceeds that of the donor. For the electron donor DIPEA, the oxidation potential is given. For consistency, it is converted to the corresponding reduction potential (\textit{i.e.} with inverted sign) when evaluating the energy balance.}    
	\label{fig:latimer}         
\end{figure}

The photochemical reduction of nickel is carried out using the upconverted annihilator perylene as a photocatalyst, diisopropylethylamine (DIPEA) as an sacrificial electron donor and \IUPAC{nickel(\Romannum{2}) chloride hexahydrate} ($\mathrm{NiCl}_2\cdot 6\mathrm{H}_2\mathrm{O}$) as a metal compound that supplies the reaction with $\mathrm{nickel}^{2+}$ cations. Starting with the perylene in the excited $\mathrm{S}_1$ state ($E_{{\text{S}_1}}\text{(per)} = \SI{2.76}{\electronvolt}$) \cite{wu2012light}, generated by sTTA-UC, the metallic $\mathrm{nickel}^{(0)}$ can be formed by two different redox pathways (Fig. \ref{fig:latimer}). To estimate, which mechanism could play a dominating role, the excited state redox potentials of perylene are considered. The excited state reduction potential of perylene in the S1 state is approximated as\cite{romero2016organic}
\begin{align}
    E(\text{per})_\text{red}^* &= E(\text{per})_\text{1/2}^\text{red} + E(\text{per})_{0,0}
\end{align}
where $E(\text{per})_\text{red}^*$ refers to the excited state reduction potential, $E(\text{per})_\text{1/2}^\text{red}$ to the ground state reduction potential of the photocatalyst and $E(\text{per})_{0,0}$ to the excited state energy approximated as $E_{\mathrm{S}_1}(\text{per})$. The perylene used in the resist exhibits ground state reduction potentials of $E_\text{red}(\text{per})=\SI{-1.67}{\volt}$ ($\text{per} + 1 e^- \ch{<=>} \text{per}^{\bullet -}$) vs. saturated calomel electrode (SCE) in DMF\cite{handbook} and $E_\text{red}(\text{per}^+)=\SI{+0.85}{\volt}$ ($\text{per}^{\bullet +} + 1 e^- \ch{<=>} \text{per}$) vs. SCE in acetonitrile\cite{handbook}. Thus, the excited state reduction potentials of perylene are calculated as $E_\text{red}(\text{per}*)=\SI{1.09}{\volt}$ ($\text{per} + 1 e^- \ch{<=>} \text{per}^{\bullet -}$) and $E_\text{red}(\text{per}^{+*})=\SI{3.61}{\volt}$ ($\text{per}^{\bullet +} + 1 e^- \ch{<=>} \text{per}$) respectively. For the reductive half reaction (Fig. \ref{fig:latimer}, right side), the excited state reduction potential of perylene ($E_\text{red}(\text{per}^*)=\SI{1.09}{\volt}$ ($\text{per} + 1 e^- \ch{<=>} \text{per}^{\bullet -}$)) is compared to the oxidation potential of the electron donor DIPEA ($E_\text{ox}(\text{DIPEA})=\SI{0.86}{\volt}$ ($\text{DIPEA}\ch{<=>}\text{DIPEA}^{\bullet +}+1e^-$) vs. SCE in DCM)\cite{goliszewska2020photoredox}, leading to the suggestion of the reaction taking place. With a reduction potential  of $E_\textbf{red}(\text{Ni}) = \SI{-0.48}{\volt}$ (\ch{Ni^{2+} + 2 e^- <=> Ni^0}) vs. SCE \cite{harris2014lehrbuch} of the nickel compound, the reduced perylene radical anion is able to reduce the nickel cations to metallic nickel. Since the oxidation state of the nickel ions from $\mathrm{NiCl}_2\cdot 6\mathrm{H}_2\mathrm{O}$ is two, two sacrificial electron donor molecules are needed to form the bulk metal \ch{Ni^{(0)}}.

For the oxidative half-reaction pathway to take place (Fig. \ref{fig:latimer}, left side), the excited state reduction potential of perylene ($E_\text{red}(\text{per}^{+*})=\SI{3.61}{\volt}$($\text{per}^{\bullet +} + 1 e^- \ch{<=>} \text{per}^{*}$)) is compared to the reduction potential of the nickel pair ($E_\text{red}(\text{Ni}) = \SI{-0.48}{\volt}$ (\ch{Ni^{2+} + 2 e^- <=> Ni^0}) vs. SCE) \cite{harris2014lehrbuch}. However, this reaction pathway is revealed to be hindered by the strongly oxidative potential of the perylene radical cation, which would be formed during this reaction. Thus, the oxidative pathway is considered as strongly suppressed. Alternatively, energy transfer from the perylene \ch{S_1} state to the $\text{Ni}^{2+}$ could be suggested, but is suppressed due to selection rules since aqueous $\text{Ni}^{2+}$ has a triplet ground state\cite{NIST_ASD}.

To further investigate the reductive pathway, the quenching mechanism was evaluated by Stern-Volmer measurements (Fig. S5). The quenching constant of $k_\text{q}^\text{DIPEA}=\SI{2.24(0.03)e9}{\per\molar \per\second}$ for perylene being quenched by DIPEA suggests a dominantly diffusion-controlled mechanism. Thus, the concentration of the electron donor DIPEA could significantly influence the efficiency of the DLW process.

A sketch of the DLW setup as well as further details can be found in a previous paper\cite{hering2016automated} and in the Methods section. The photoresist is freshly prepared before each application to ensure that its composition remained unchanged. Details of the photoresist can be found in the Methods section.
For application in the DLW system, a glass substrate is sputtered with a \SI{3}{\nano\meter} thin transparent layer of iridium. Iridium has strong spin-orbit coupling due to heavy atom effect and thus enhances the intersystem-crossing to the triplet states required for sTTA-UC\cite{Lakowicz2006}. The photoresist is sandwiched between the coated substrate and a glass window using tape as a spacer. To avoid evaporation of the resist the cell is sealed using Marabu Fixogum, a commercially available adhesive (for further details see the Methods section). 

The power of the laser is adjusted to provide \SI{15}{\milli\watt} power at the entrance pupil of the microscope objective at \SI{100}{\percent} printing power. For most printing processes, the power is tuned down to \SIrange[]{60}{80}{\percent}. The scan speed for printing of the structures in Fig. \ref{fig:SEM_and_EDX} is \SI{100}{\micro\meter\per\second}. We observe a slight degradation of the printing efficiency after \SI{60}{\minute}, which is most probably due to consumption of $\mathrm{Ni}^{2+}$ ions.

After exposure the substrate is dipped into acetone for \SI{2}{\minute} and isopropyl alcohol for \SI{1}{\min} to dissolve the remaining resist and reveal the printed structures. Finally, the substrate is dried using a flow of gaseous nitrogen.

\subsection{Direct Laser Written Nickel structures and their properties}

\begin{figure}             
	\centering                
	\def\svgwidth{250pt}    
	\includegraphics[width = 0.47\textwidth]{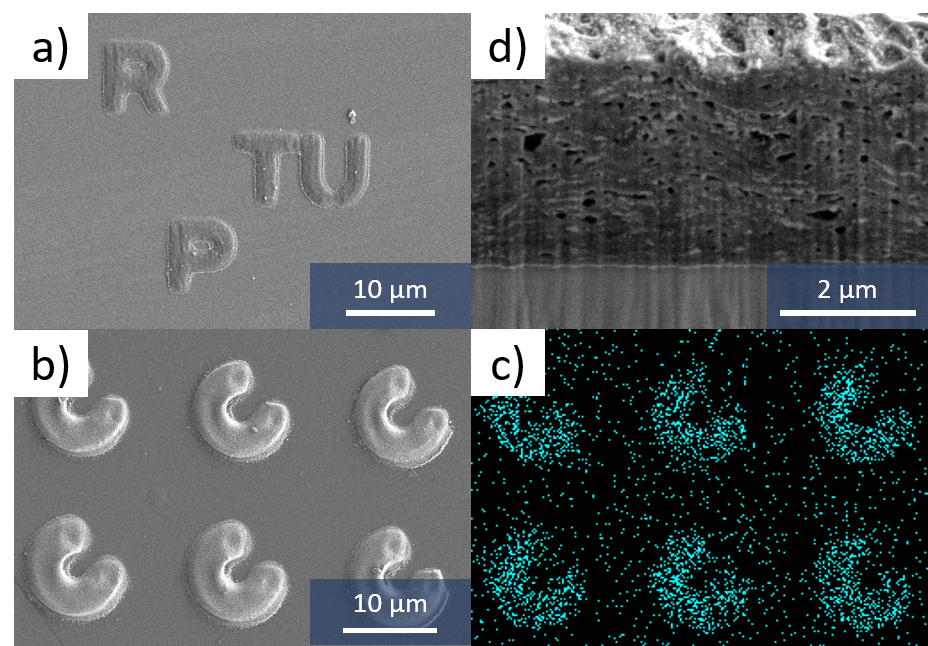}
	\caption{\small SEM images of 2.5D nickel structures printed using the proposed resist. a) Our university's logo. b) A close up of an array of ring segments. c) EDX obtained countmap visualizing the distribution of nickel matched to the sample area in b). d) Cross section of a structure generated using a focused ion beam.}    
	\label{fig:SEM_and_EDX}         
\end{figure}

As a proof of concept, we print 2.5D nickel structures. Fig. \ref{fig:SEM_and_EDX} a) and b) show exemplary SEM images of printed structures. Fig. \ref{fig:SEM_and_EDX} c) displays the spatial distribution of nickel obtained by energy dispersive X-ray spectroscopy (EDX), matching the area displayed in Fig. \ref{fig:SEM_and_EDX} b). Is is clearly recognizable that nickel is detected where the printed material is deposited. Besides some less densely distributed carbon, most likely resulting from heat generation in the laser focus, all other elements that could be detected by EDX are detected equally over the whole area, or even more at the area without the structures, and thus appear to be components of the substrate. All countmaps of elements that EDX was able to detect in the sample are shown in Fig. S6.

To determine the material density of the printed structures, cross-sections from a printed structure are prepared using a focused ion beam to sequentially slice the sample. An SEM image is taken for each slice during the cross sectioning. Fig. \ref{fig:SEM_and_EDX} d) shows one of the SEM images with a side view onto the surface, on which pores with a diameter of about \SIrange{50}{200}{\nano\meter} are visible. The images were analyzed with a graphics processing software. The density of the printed material is found to be \SI{96(3)}{\percent}.

\subsection{Ferromagnetic characteristics of the printed structures}

\begin{figure*}[ht]     
	\centering     
	\fontsize{7pt}{12pt}\selectfont            
	\def\svgwidth{280pt}    
 	\includegraphics[width = \textwidth]{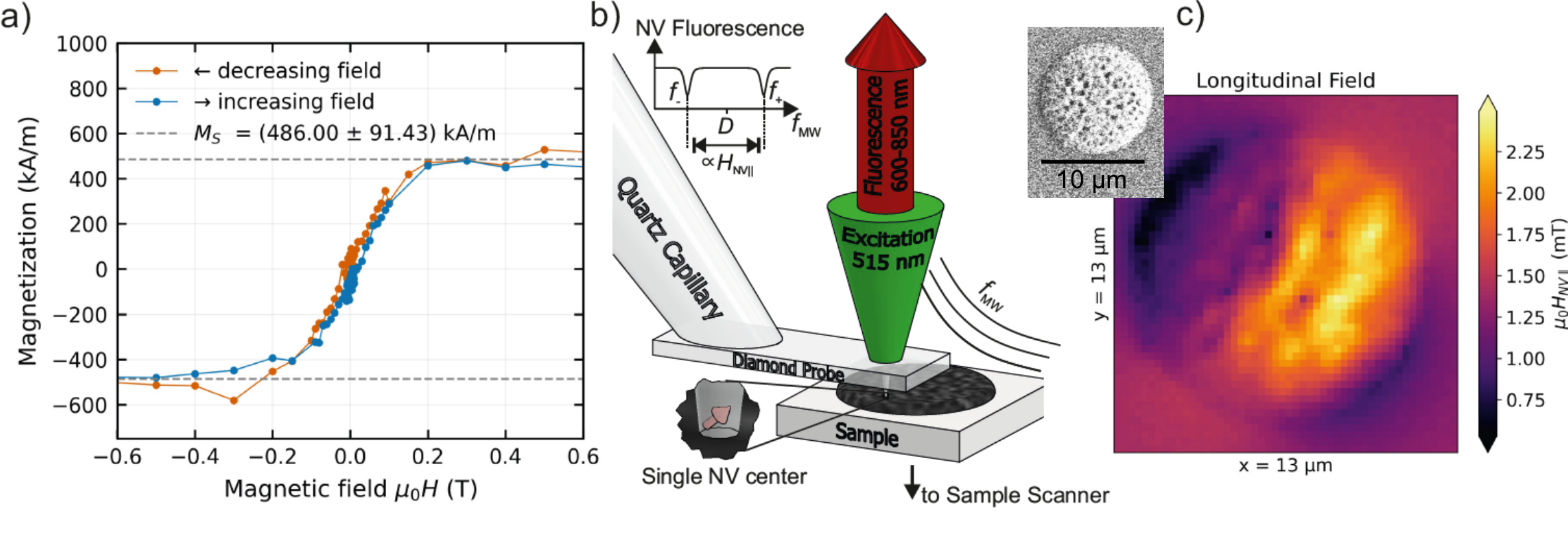}
	\caption{\small Magnetic characterization of nickel microdots. a) Room-temperature magnetic hysteresis loop of a nickel dot array of elements analogous to the structure shown in c. b) Schematic of the scanning NV magnetometer utilized for single-structure characterization. The upper-left inset schematically displays an optically detected magnetic resonance (ODMR) spectrum, where the resonances $f_-$ ($m_\mathrm{S}=0 \leftrightarrow m_\mathrm{S}=-1$) and $f_+$ ($m_\mathrm{S}=0 \leftrightarrow m_\mathrm{S}=+1$) of the NV electronic ground state are visible. Within the small-field approximation, the Zeeman splitting around the zero-field splitting $D$ is proportional to the magnetic field projection onto the NV quantization axis, $\mu_0H_\mathrm{NV||}$. c) Top left: SEM image of the investigated microdot. The main panel shows the quantitative stray field map ($\mu_0H_\mathrm{NV||}$) recorded scanning the NV at a height of \SI{500}{\nano\meter} above the nickel structure. The structure was saturated at \SI{300}{\milli\tesla} and measured in remanence, using a small bias field to ensure clear splitting of the NV resonances.}
\label{fig:magnetic_characterization}         
\end{figure*}

To evaluate the magnetic properties of the printed nickel structures, we first performed vibrating sample magnetometry on an array of microdots (diameter $\approx$ \SI{10}{\micro\meter}, height $\approx$ \SI{0.6}{\micro\meter}).  
The magnetic moment was recorded for both decreasing (Direction 1, orange) and increasing (Direction 2, blue) external fields $\mu_0H_\mathrm{ext}$ after initial saturation at $\pm$\SI{9}{\tesla}, respectively. We calculated the magnetization from the measured moment and sample volume obtained from confocal microscopy as described in the Methods section. 
Fig. \ref{fig:magnetic_characterization}a) displays the corresponding hysteresis loop. 
The obtained saturation magnetization of $M_\mathrm{S} = \qty{486\pm91}{\kilo\ampere\per\meter}$ agrees with the literature value of \qty{480}{\kilo\ampere\per\meter} \cite{sulitanu_saturation_1992}. The uncertainty of our value $M_\mathrm{S}$ is primarily dominated by the determination of the microdot array volume via confocal microscopy.
The resulting loop exhibits a saturation field of approximately \SI{200}{\milli\tesla} and a remanence of about \SI{11}{\percent} of $M_\mathrm{S}$. 
These values are significantly higher than those of bulk nickel, which typically saturates at fields below \qty{1}{\milli\tesla} and exhibits negligible remanence at room temperature \cite{Bozorth1951}. 
While a direct comparison to bulk material neglects the demagnetizing effects inherent to microstructures, micromagnetic simulations for an ideal cylindrical nickel dot (\qty{10}{\micro\meter} diameter, \qty{0.6}{\micro\meter} height) predict a saturation field ($M \geq 0.95 \cdot M_\mathrm{S}$) of only \qty{50}{\milli\tesla}.
This discrepancy suggests pronounced domain wall pinning, likely originating from the internal structure of the printed material, specifically the gaps and cavities visible in the cross-section in Fig.~\ref{fig:SEM_and_EDX}c).

\subsection{Nitrogen-vacancy magnetometry}

To investigate the magnetic properties of individual structures, we employ a commercial scanning NV setup (ProteusQ, Qnami, see Methods section for further details), allowing to quantitatively measure the magnetic stray field above the sample (Fig. \ref{fig:magnetic_characterization}b). The measured nickel disc is prepared in its remanent state. The sample was saturated with an in-plane field of approximately \SI{300}{\milli\tesla} and the NV probe was scanned at a lift height of \SI{500}{\nano\meter}, using only a weak bias field to spectrally separate the NV spin resonances. The resulting map of the longitudinal field projection $\mu_0 H_{\mathrm{NV}\parallel}$ is shown in Fig. \ref{fig:magnetic_characterization}c), revealing pronounced stray‑field contrast and ferromagnetic ordering of the printed nickel. 
The stray-field modulation spans the entire microdot without significant internal fluctuations, indicating that the detected dipolar field originates from the macroscopic remanent magnetization of the structure. 
By comparing the experimental data with the calculated dipolar field of a fully saturated dot ($\approx$ \qty{60}{\milli\tesla}), we estimate the remanence of this specific structure to be \qtyrange{1}{2}{\%} (see Fig. S7). This value is notably lower than the \qty{11}{\%} average remanence determined \textit{via} VSM for the microdot array.

The absence of resolvable magnetic domain contrast in the stray-field image can be attributed to two factors. First, a characteristic domain size below \qty{500}{\nano\meter} would lead to a rapid decay of the field modulation with increasing distance from the surface. Consequently, such domains would remain unresolved at the given probe-to-sample distance. The formation of small domains is likely promoted by the internal morphology of the material, where cavities act as pinning sites or locally break the exchange coupling—the fundamental interaction that maintains long-range ferromagnetic alignment—thereby favoring a reduction in the characteristic magnetic length scales.

Second, a layered growth process may favor an antiferromagnetic-like coupling between consecutive layers, as the aforementioned cavities predominantly reside at the interfaces between printed layers. These may partially decouple the magnetic moments of adjacent layers whereas intra-layer exchange coupling remains largely undisturbed. This could result in a lamellar structure with compensated magnetization directions, leading to a high degree of magnetic flux closure within the dot.

\subsection{Conclusions}
A new photochemical strategy for direct laser writing of metallic material was introduced, addressing key limitations in the additive manufacturing of ferromagnetic metals. By combining photochemical deoxygenation, sensitized triplet–triplet annihilation upconversion, and photoredox-catalyzed metal reduction, a stable DLW process for Nickel was realized under ambient conditions using a continuous wave \SI{532}{\nano\meter} excitation. 

Nickel structures were fabricated with the proposed photoresist as proof of concept, exhibiting high material density and clear ferromagnetic order, as evidenced by both macroscopic VSM and high-resolution scanning NV magnetometry.
While the measured saturation magnetization is consistent with bulk nickel, the observed saturation fields ($\approx \qty{200}{\milli\tesla}$) and remanence ($\approx \qty{11}{\%}$ in VSM measurements) point toward pronounced domain-wall pinning. Furthermore, the absence of significant magnetic domain contrast at sub-micrometer distances during the scanning NV magnetometry characterization in the structure's remanent state suggests a fragmented domain state with a high degree of magnetic flux closure.

This work demonstrates that sTTA-UC-driven photochemistry is a powerful and versatile tool for additive manufacturing of functional metallic microstructures. The presented concept provides a promising route towards complex magnetic microcomponents for potential applications in microrobotics, sensing, and integrated microsystems.

\section{Methods} 

\subsection{Photophysical characterization}

Absorption spectroscopy was carried out with a JASCO V-780 double-beam UV/vis to NIR spectrophotometer. For measurements up to \SI{350}{\nano\meter} a deuterium lamp and from \SIrange{350}{800}{\nano\meter} a halogen lamp was facilitated. The luminescence spectra were recorded with a HORIBA FluoroLog3 22$\tau$ with a \SI{450}{\watt} Xenon lamp and double-grating monochromator units for the excitation light and emitted light by the sample. Further, luminescence lifetimes were measured by time-correlated single photon counting (TCSPC) with the HORIBA DeltaFlex using a NanoLED (pulse width < \SI{1.3}{\nano\second}) at a wavelength of \SI{389}{\nano\meter}. The internal response function was recorded using suspension of $\text{LUDOX}^\text{\textregistered}$ in water to scatter the excitation light. 

\subsection{Further data evaluation}

The  recorded TCSPC decays were analyzed using the HORIBA DecayAnalysis software. All data were fitted using reconvolution fitting with the internal response function and a mono exponential function (Fig. S2 \& S3).
For evaluation of the Stern-Volmer constant $K_\mathrm{SV}$ and the quenching rate of perylene with DIPEA, the quencher concentration $[\mathrm{Q}]$ was plotted against the excited state lifetime $\tau$ of perylene with DIPEA and $\tau_0$ without added quencher. The data were fitted according to the equation \ref{equation: stern volmer fit}.
\begin{align} \label{equation: stern volmer fit}
    \frac{\tau}{\tau_0} = 1 + K_\mathrm{SV}
\end{align}
To extract the quenching rate $k_\mathrm{q}$ the relation \ref{equation: quenching rate} was used.
\begin{align} \label{equation: quenching rate}
    K_\mathrm{SV} = k_\mathrm{q} \cdot \tau_0
\end{align}

\subsection{Preparation of the photoresist}

The resist is prepared under ambient conditions. The final resist contains \SI{10}{\milli\molar} perylene, \SI{100}{\micro\molar} Erythrosine B, \SI{1}{\milli\molar} $\mathrm{NiCl}_2\cdot 6\mathrm{H}_2\mathrm{O}$ and \SI{574}{\milli\molar} DIPEA in DMI. 

DMI (>99\%) and perylene (>98\%) were purchased from Tokyo Chemical Industry Co., Ltd. (TCI, Tokyo, Japan). $\mathrm{NiCl}_2\cdot 6\mathrm{H}_2\mathrm{O}$ (>98\%) was purchased from VWR International, LLC (Radnor, PA, USA). DIPEA (>99\% for synthesis) was purchased from Carl Roth GmbH + Co. KG (Karlsruhe, Germany) and Erythrosine B was purchased from ClinTech Limited (Manchester, UK). All chemicals are used as bought.

\subsection{DLW setup}

For DLW, a self built setup is used. The laser is a continuous wave diode-pumped solid-state Neodymium Vanadate (Nd:$\mathrm{YVO}_4$) with a wavelength of \SI{532}{\nano\meter} and an average power of up to \SI{2.2}{\watt} (Coherent Verdi V-2). The laser light is focused onto the substrate with a NA 1.4 oil immersion objective. Scanning of the laser focus is enabled by galvanometric mirrors. The laser power can be adjusted using an accousto-optical modulator. A sketch of the setup as well as further details can be found in a previous paper\cite{hering2016automated}.

To test the theory if the proposed process could be assisted by additionally trapping generated particles by enhancement of their near field \textit{via} localized surface plasmon resonance induced by 2PA like Wang \textit{et al.} did \cite{wang2024free}, a \SI{780}{\nm} laser with \SI{150}{\femto\second} pulses and a repetition rate of \SI{80}{\mega\hertz} with an average power of up to \SI{1.5}{\watt} was added to the setup. The laser foci of both lasers were aligned in the printing process. The structures printed with this setup did not show any noticeable improvement compared to the ones without the added pulsed laser.

Following a similar approach, experiments with a pulsed laser that emits a wavelength that is twice the excitation wavelength of the sensitizer were carried out to investigate the mechanism of employing a 2PA process to sensitize the sTTA-UC process. For this purpose, the compression module of a Primus series laser by Stuttgart Instruments with a fixed wavelength at \SI{1040}{\nano\meter} with an average output power of \SI{1}{\watt} at a pulse duration of \SI{450}{\femto\second} and a repetition rate of \SI{40}{\mega\hertz} was employed and focused with a NA 0.5 objective into a vial containing \SI{2}{\milli\liter} of the resist. The laser power corresponds to a peak power of $\sim \SI{5.5e4}{\watt}$. While a yellow luminescence consistent with the fluorescence of Erythrosine B at \SI{553}{\nano\meter}\cite{cheng2018erythrosin} was observed, no observable metal particles were formed.

\subsection{Determination of the ferromagnetic characteristics of the printed structures}
Vibrating sample magnetoetry (VSM) measurements were conducted using a commercial setup (PPMS DynaCool, Quantum Design) at a temperature of \qty{300}{\kelvin}. The sample was glued to a quartz sample holder that oscillates inside of a pick-up coil at \qty{40}{\hertz} and with an amplitude of \qty{2}{\milli\meter}. The induced voltage in the pick-up coil is used to determine the magnetic moment of the sample as a function of externally applied magnetic field. The magnetic field is applied via a superconducting solenoid coil around the sample and pick-up coil. 

For each magnetic field setting, the average of three measurements was used to precisely determine the magnetic moment, whereas only clusters with a standard deviation of less than \qty{2}{\nano\ampere\meter^2} were considered in the evaluation. 

The volume of the sample used in the VSM measurement was obtained using a confocal microscope. After measuring the topography of the sample, the volume was evaluated using a Matlab script. By tweaking the parameter settings in the script based on the possible factors, the volume was estimated both upwards and downwards. The volume was found to be \SI{1.05(0.17)e-14}{\cubic\meter}. With the determined material density of \SI{96(3)}{\percent}, the magnetic volume of the sample is given by \SI{1.01(0.19)e-14}{\cubic\meter}.
Dividing the magnetic moment by this volume yields the magnetization.

A linear diamagnetic background, originating mainly from the quartz sample holder and substrate, was removed by calculating the mean slope and offset from linear fits within the saturated regimes \SI{0.3}{\tesla}~$< |\mu_0H_\mathrm{ext}| <$~\SI{1}{\tesla}. This regime was also used to calculate the saturation magnetization by averaging across the magnetic moments of all data points within this region.

\subsection{Nitrogen-vacancy magnetometry}

To investigate the magnetic properties of single structures, we employ Nitrogen-vacancy magnetometry. Nitrogen‑vacancy (NV) centers in diamond are point defects consisting of a substitutional nitrogen atom adjacent to a carbon vacancy. Their electronic ground state is a spin triplet that can be initialized and read out optically and is coherently manipulated with microwaves, which enables quantitative magnetometry under ambient conditions \cite{Rondin_2014}. In an external magnetic field, the degeneracy of the $m_s=\pm 1$ spin states is lifted and the corresponding optically detected magnetic resonance (ODMR) frequencies $f_-$ and $f_+$ are Zeeman shifted. From the splitting between these resonances, the projection of the local magnetic field onto the NV quantization axis, $\mu_0 H_{\mathrm{NV}\parallel}$, can be extracted using the NV gyromagnetic ratio, providing a direct measure of the stray field above the sample. 

For the measurements, a commercial scanning NV setup (ProteusQ, Qnami) is operated in frequency modulated (FM) atomic force microscopy (AFM) mode, with a single NV center at the apex of a diamond tip (MX+ series, (100)-oriented). The NV center is excited with a \SI{515}{\nano\meter} laser focused onto the tip through a microscope objective (NA = 0.7), and the resulting fluorescence is detected while applying a microwave field of frequency $f_\mathrm{MW}$ to record the ODMR spectra (Fig. \ref{fig:magnetic_characterization}b).

\begin{acknowledgement}

This work was funded by the Deutsche Forschungsgemeinschaft
(DFG, German Research Foundation) under project number TRR 173–268565370, Spin+X (Project A12) and project number 172116086 - SFB 926 (project B11).

One of the laser systems (Stuttgart Instruments) used for this study has been funded by the Deutsche Forschungsgemeinschaft (DFG, German Research Foundation) – project number 459322035.

We acknowledge the use of our Scanning NV Magnetometer, funded by the Deutsche Forschungsgemeinschaft (DFG, German Research Foundation) - 491229782 within the major instrumentation initiative “Spin-based quantum light microscopy (SQLM)”.

We acknowledge support from nano structuring center (NSC) at the RPTU and thank especially Dr. Thomas Löber for preparing the focused ion beam cross-sections.

K. Rediger gratefully acknowledges the financial support by the Fonds der Chemischen Industrie (FCI, 115571) in the form of a Kekulé scholarship and thanks Prof. Dr. Maria Wächtler (Christian-Albrechts-Universität zu Kiel) for her continuous and fruitful support.

\end{acknowledgement}

\begin{suppinfo}


The following files are available free of charge.
\begin{itemize}
  \item Supplementary information: A PDF containing figures with more detailed measurement data (spectroscopic data of the resist, EDX obtained countmaps of a nickel structure) and simulated data (modeling of the magnetic stray field for a nickel structure).
\end{itemize}

\end{suppinfo}

\bibliography{quellen}

\end{document}